\documentclass[12pt,manuscript]{aastex}

\usepackage{txfonts}

\usepackage{graphicx}

\usepackage{longtable} 

\usepackage{natbib}

\slugcomment{For submission to the Astrophysical Journal}
\shorttitle{Activity Patterns in a Model Active Region}
\shortauthors{Bradshaw and Viall}
 
\begin{document}
\tighten
\title{Patterns of Activity in a Global Model of a Solar Active Region}
 
\author{S. J. Bradshaw}
\affil{Department of Physics and Astronomy, Rice University, Houston, TX 77005, USA}
\email{stephen.bradshaw@rice.edu}\and

\author{N. M. Viall}
\affil{NASA Goddard Space Flight Center, Greenbelt, MD 20771, USA}
\email{Nicholeen.M.Viall@nasa.gov}
 

\begin{abstract}
In this work we investigate the global activity patterns predicted from a model active region heated by distributions of nanoflares that have a range of frequencies. What differs is the average frequency of the distributions. The activity patterns are manifested in time lag maps of narrow-band instrument channel pairs. We combine hydrodynamic and forward modeling codes with a magnetic field extrapolation to create a model active region and apply the time lag method to synthetic observations. Our aim is not to reproduce a particular set of observations in detail, but to recover some typical properties and patterns observed in active regions. Our key findings are the following. 1. cooling dominates the time lag signature and the time lags between the channel pairs are generally consistent with observed values. 2. shorter coronal loops in the core cool more quickly than longer loops at the periphery. 3. all channel pairs show zero time lag when the line-of-sight passes through coronal loop foot-points. 4. there is strong evidence that plasma must be re-energized on a time scale comparable to the cooling timescale to reproduce the observed coronal activity, but it is likely that a relatively broad spectrum of heating frequencies are operating across active regions. 5. due to their highly dynamic nature, we find nanoflare trains produce zero time lags along entire flux tubes in our model active region that are seen between the same channel pairs in observed active regions.
\end{abstract}

 

\section{Introduction}
\label{introduction}

In the work presented here we focus on understanding the global properties of heating in solar active regions. Current ideas concerning the manner in which the non-flaring solar corona is supplied with plasma heated to temperatures of several million degrees hold that small-scale, impulsive events called "nanoflares" may hold the key to unravelling this mystery. The precise physical mechanism underlying these events is not presently known, but it may be related to the reconnection of tangled and twisted magnetic field lines driven by the motions of their foot-points at the solar surface \citep{Parker1988} and/or by the turbulent dissipation of counter-propagating Alfv\'{e}n waves reflected as they encounter the changing refractive index of the gravitationally stratified atmosphere \citep{Asgari-Targhi2013}. 

Modern solar observatories can help to elucidate several important nanoflare properties and when combined with numerical models, very strong constraints can be placed on the region of the parameter space they must occupy. Of particular interest are the amount of energy released in a single heating event (the canonical nanoflare deposits $\sim 10^{24}$~erg), the duration of each event (a few seconds to several minutes), the spatial extent over which the energy is released (perhaps a few hundred to a thousand km), the heating location, and the frequency at which plasma is re-energized. Note that in a series of recent papers, \cite{Klimchuk2012,Klimchuk2014}; and \cite{Bradshaw2015} have ruled out chromospheric nanoflares as the dominant source of coronal plasma. They showed through analytical and modeling work that chromospheric nanoflares are not consistent with observed spectral line profiles, and the energy release site must lie above the chromosphere. However, they do not preclude chromospheric nanoflares as a source of energy for the chromosphere itself. Other questions concern the sympathetic nature of nanoflares. Are they spatially and temporally isolated events? Are many such events triggered on adjacent, sub-resolution magnetic strands within a finite time-window? Do nanoflare storms and/or several reconnections to the relaxed state comprise a single event \citep{Klimchuk2009,Viall2011,LopezFuentes2010,LopezFuentes2015}? For a recent review of the most important aspects of coronal heating that need to be understood to solve the problem see \cite{Klimchuk2015}.

The property of interest that we directly address in this paper is the ``recharging" timescale of the nanoflare mechanism, governing the frequency at which plasma on any given magnetic strand is re-energized. This has given rise to the terms {\it low-} and {\it high-frequency nanoflare trains} in the literature \citep{MuluMoore2011,Warren2011,Bradshaw2012,Reep2013a}. Low frequency nanoflares are defined by inter-event periods that are longer than a characteristic cooling timescale, which allows the coronal plasma on a magnetic strand to cool and drain to sub-million degree temperatures and low densities before the next nanoflare is triggered. In this case, a broad range of temperatures can be observed. Conversely, high frequency nanoflares occur with inter-event periods that are significantly less than a cooling timescale, the lower limit of which yields effectively steady heating, and maintain the corona with relatively hot and dense plasma. It is not yet known whether the inter-event period is in fact {\it intermittent} or {\it periodic}. One idea explored by \cite{Cargill2014} is that the inter-event period is proportional to the energy of the next event in the train. One physical interpretation of this idea in the reconnection scenario is that the magnetic field becomes more tangled and twisted over a longer time period, and eventually releases more energy.

The emission measure, $EM(T)=\int n^2 ds$, which quantifies the amount of plasma as a function of temperature along the line-of-sight, has proven useful in terms of connecting the properties of the underlying heating mechanism to particular properties of this important coronal observable. For example, the magnitude and the temperature $(T_p)$ of the peak $EM$ provide a means to estimate the input energy required to power the observed radiation. Denser, hotter coronal plasma naturally requires more energy to ablate and heat sufficient plasma from lower-lying layers of the atmosphere; simple scaling laws \citep[e.g.][]{Rosner1978} or numerical models \citep[for greater accuracy, e.g.][]{Klimchuk2008,Cargill2012} can be used in tandem with the $EM$ to calculate the energy requirements. \cite{Cargill1994,Cargill1995} predicted that the $EM$ associated with low frequency nanoflare heating in active regions should peak at $2 \leq T_p \leq 4$~MK \citep[and requires impulsively heated magnetic strands to exceed 5~MK before cooling, to reach the observed over-densities at 1~MK:][]{Aschwanden2000,Spadaro2003}. The $EM$ was also predicted to scale as $EM(T) \propto T^\alpha~(\alpha \approx 2)$ below $T_p$. Observations show that $2 \leq \alpha \leq 5$ \citep[Table~3 of][]{Bradshaw2012}, which is connected to the inter-event period of the nanoflares comprising the train. Shallow slopes (small $\alpha$) imply low frequency heating and that the plasma cools through a broad temperature range before it is re-energized. Steep slopes (large $\alpha$) imply higher frequency heating and the plasma temperature is maintained within a narrower range.

Above $T_p$, $EM(T) \propto T^{-\beta}~(\beta \gg 1)$. This implies that the bulk of the coronal plasma has temperature $T \leq T_p$ with just a small amount of hot plasma at $T > T_p$, suggesting that the associated emission will be relatively faint and hard to detect. \cite{Warren2012} surveyed 15 active regions and produced $EM$ plots for selected inter-moss regions (the region between loop foot-points) in each; they found $6.1 \leq \beta \leq 10.3$. Nonetheless, evidence for super-heated ($T\sim10$~MK) plasma in non-flaring active regions has been gradually accumulating \citep{McTiernan2009,Schmelz2009a,Schmelz2009b,Reale2009a,Reale2009b,Testa2011,Miceli2012,Testa2012,Brosius2014,Petralia2014a,Caspi2015}. The presence of a hot component to the $EM$ is also indicative of the nanoflare frequency. In the case of low frequency heating, the energy per particle deposited during a heating event may be sufficient to (briefly) produce high temperature plasma, following significant cooling and draining of the corona. In the case of high frequency heating, the plasma density and temperature is maintained within a much narrower range, and substantial emission at a significantly higher temperature would require an anomalously energetic event.

$EM$ distributions primarily depend upon spectroscopic data, for which fast rastering requires limited emission lines and incomplete spatial information, rendering them most useful for examining spatially localized properties of the heating mechanism. We are interested in understanding the global properties of active region heating and must examine the evolution of the observed emission across the whole region, which necessitates gathering a large amount of data at a high cadence. Consequently, an alternative diagnostic for global properties of active region heating is needed. In this case, high-resolution, imaging data can be well-suited to our purposes. \cite{Viall2012} developed a tool to produce time lag maps by correlating light-curves in pairs of Solar Dynamics Observatory (SDO) Atmospheric Imaging Assembly (AIA) \citep{Lemen2012} channels for every spatial pixel of an image. Each pixel is colored according to the channel of the pair in which the emission first appears and the amount of time, or time lag, before it appears in the other channel. Since each channel is sensitive to a unique temperature range, the maps give an immediate picture of the global pattern of heating and cooling in the active region. The full-Sun FOV, high-resolution (0.6\arcsec) and fast cadence (12~s) of AIA make that instrument ideally suited to this task. With the six channels of AIA, the time lag method results in 15 different maps of time lags between pairs of channels, which is an extremely strong set of constraints for nanoflare models.

In the following Sections we present a series of results obtained by applying the time lag method to a set of synthetic AIA active region observations calculated for statistically intermediate- and statistically high-frequency nanoflare heating, and a control experiment in which the signal is expected to be dominated purely by post-nanoflare cooling. Based on our findings for all 15 time lag maps, we demonstrate that time lag maps derived from observational data are consistent with coronal nanoflares, and particular differences between the time lag maps for key channel pairs can be used to distinguish between intermediate- and high-frequency heating. In Section~\ref{model} we describe the field-aligned hydrodynamic and forward modeling used to predict the photon counts measured in the AIA channels for each of the nanoflare heating scenarios explored, and the 3D active region visualization technique used to render the synthetic observations. In Section~\ref{results} we discuss our findings from this investigation in detail, and in Section~\ref{SandC} provide a summary and a set of conclusions.

\section{Numerical Model}
\label{model}

\subsection{Field-Aligned Hydrodynamics and Forward Modeling}

The low plasma $\beta$ in the upper solar atmosphere and the inhibition of cross-field mass and energy transport mean that we can treat each magnetic field line as an isolated atmosphere, solving for the plasma structure and evolution in the field-aligned direction as it responds to an energy release. The system of equations appropriate to this treatment are solved using HYDRAD: the HYDrodynamics and RADiative emission model \citep{Bradshaw2013}. HYDRAD solves the time-dependent equations for the evolution of mass, momentum, and energy for multi-fluid plasma (electrons, ions and neutrals) in arbitrary magnetic geometries taking account of the field-aligned gravitational acceleration, and includes bulk transport (with shock capturing), thermal conduction (with heat flux limiting and delocalization), viscous interactions, gravitational energy, Coulomb collisions, and optically-thick radiation in the lower atmosphere transitioning to optically-thin radiation (lines and continuum) in the overlying atmosphere. The heating term can be specified in a flexible, parameterized form, or in a tabulated form input from a code specifically designed to study the energy release mechanism \citep[e.g.][]{Buchlin2007}.

HYDRAD features adaptive mesh refinement that is capable of very fine resolution (to meter scales on modern hardware) and makes the code very efficient by ensuring that the spatial density of grid cells increases wherever needed within the computational domain, while keeping the total number of cells manageable. It is particularly important to resolve the steep transition region temperature gradients so that the corona is not under-filled during heating \citep{Bradshaw2013}.

The forward modeling component of this work is carried out with a powerful spectral synthesis code described in \cite{Bradshaw2011a,Bradshaw2011b,Bradshaw2012,Reep2013a}; and \cite{Reep2013b}. The output from HYDRAD provides the field-aligned evolution of the plasma in terms of the species temperatures, pressures and number densities, the bulk flow velocity, and the ion populations. The forward modeling code uses these quantities to synthesize spectra and wavelength-integrated emission from each grid cell along the field-line. The code uses atomic data provided by version 7.1 of the Chianti database \citep{Dere1997,Landi2013} and includes data for hundreds of ions and thousands of spectral lines. We note that in the present work all calculations are performed assuming ionization equilibrium. The codes are all capable of non-equilibrium ionization calculations, which we plan to address in future investigations. For the purposes of this study we are interested in general trends and are satisfied that equilibrium ionization provides a useful starting point because the dominant signal in time lag maps is from cooling plasma, where departures from equilibrium may not be substantial (though perhaps not negligible either).

\subsection{Active Region Modeling, Visualization, and Synthetic Observations}

\begin{figure}
\includegraphics[width=0.5\textwidth]{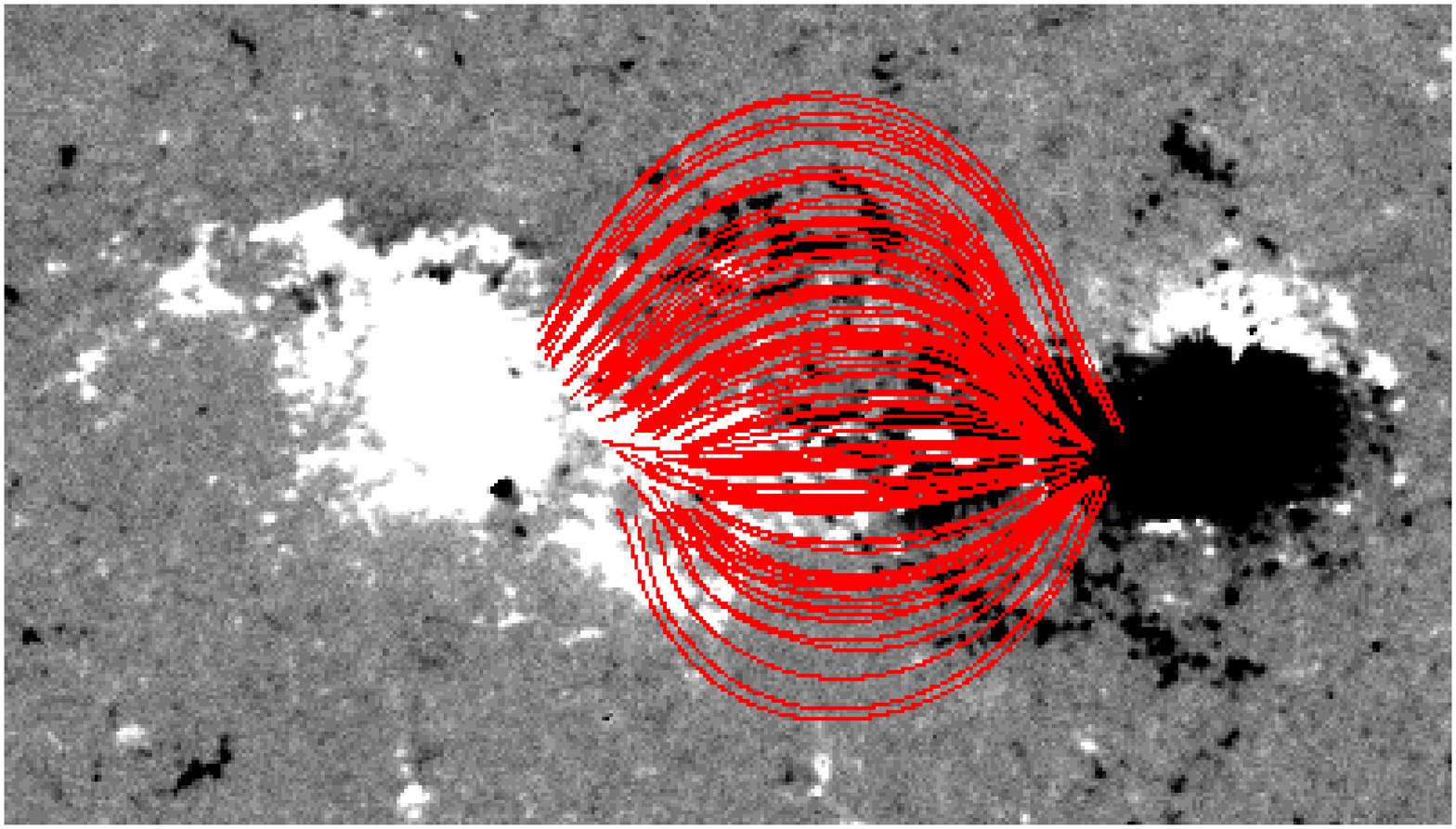}
\includegraphics[width=0.5\textwidth]{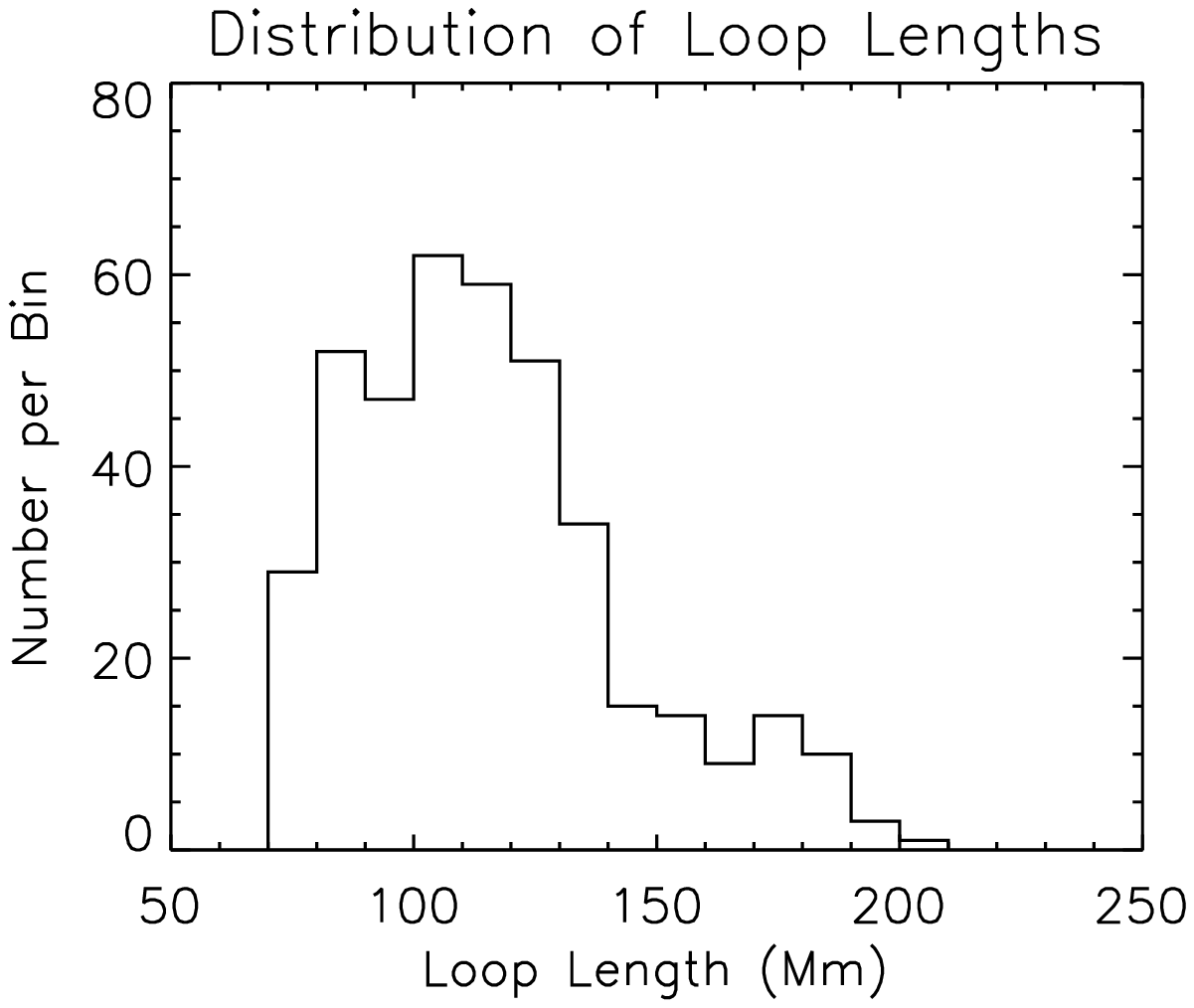}
\caption{Left-hand panel: the SDO/HMI line-of-sight magnetogram data for NOAA active region \#11640 and a selection of core magnetic field-lines extrapolated using the potential field source surface approximation. Right-hand panel: a histogram showing the distribution of extrapolated field-line/loop lengths.}
\label{fig1}
\end{figure}

A model active region core was constructed by extrapolating field lines from a SDO/HMI active region magnetogram, using a Potential Field Source Surface \citep[PFSS:][]{Schatten1969} model, to build its 3D magnetic skeleton. NOAA active region \#11640 was selected for this purpose and is shown in the left-hand panel of Figure~\ref{fig1}, with a subset ($\sim100$) of the extrapolated field-lines overlaid. This active region shows a reasonably simple dipole configuration and was located at $\sim$ zero solar longitude at 00:04 UT on 2013-01-01. Though active region fields are unlikely to be potential \citep[particularly young, newly developed active regions:][]{Wiegelmann2005}, our primary requirement is to gather a large sample of field line lengths and geometries, rather than (for example) to recover a particular topology that leads to eruptive phenomena. Furthermore, other extrapolation schemes of increasing sophistication have their own difficulties \citep[e.g.][]{DeRosa2009}. 

The PFSS software provided with SolarSoft returns a user-specified number of randomly extrapolated, closed field lines. A subset of 400 field lines with a maximum length of 250~Mm were then selected for the 3D magnetic skeleton (distribution of lengths shown in the right-hand panel of Figure~\ref{fig1}). These field lines can intersect one another along the line-of-sight, just as is likely to be the case for the real Sun. The field-aligned gravitational acceleration was calculated as a function of position for each field line (then stored as a look-up table) and the equations of force balance and energy were integrated from foot-point to foot-point, yielding hydrostatic coronal loop initial conditions that take account of the field line geometry. The loop cross-section was assumed to be constant from foot-point to apex. One could, in principle, extract the area variation from the magnetic field data and include it as an area factor in the fluid equations, but without good coronal magnetic field measurements it is a difficult additional parameter to satisfactorily constrain. The background heating was assumed to be uniform and the initial peak temperature kept well below 1~MK. A chromosphere of depth 5~Mm and isothermal temperature 0.02~MK was added to each foot-point as a source/sink of coronal mass.

\begin{figure}
\includegraphics[width=0.5\textwidth]{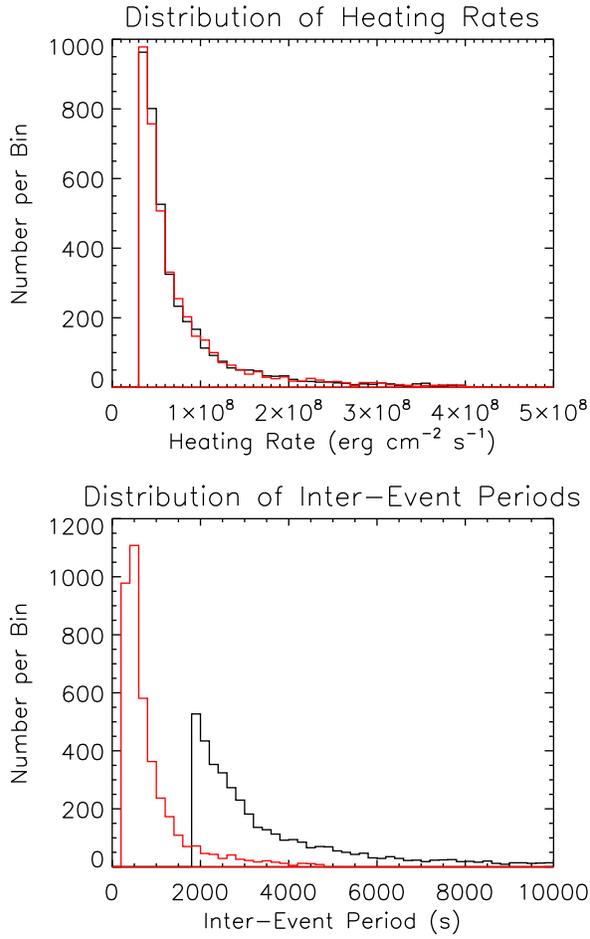}
\caption{Upper panel: the distribution of heating rates chosen for the numerical experiments. Lower panel: the distribution of inter-event periods for the numerical experiments. The black curves correspond to the intermediate frequency experiment and the red curves to the high frequency experiment.}
\label{fig2}
\end{figure}

A separate instance of HYDRAD was run for each of the 400 field lines and three numerical experiments were conducted. In the first (control) experiment, a single nanoflare was triggered on all field lines at $t=0$~s and the plasma allowed to cool thereafter, with no further events. In the second experiment, the plasma on each field line was energized by a statistically intermediate frequency nanoflare train. In the third experiment, a statistically high frequency train was used to energize the plasma. Note that these distributions of nanoflares have a range of frequencies, and what differs is the average frequency of the distributions. The heating rates were drawn from a power-law distribution of slope -2.5 \citep[-2 is the boundary between large/small events dominating the heating:][]{Hudson1991}, the same as the universal power-law found by \cite{LopezFuentes2015}, and extending two decades in energy: $10^7 \leq E \leq 10^9$~erg~cm$^{-2}$~s$^{-1}$. The upper panel of Figure~\ref{fig2} shows the nearly identical distribution of heating rates for the intermediate and high frequency experiments. The control experiment draws the heating rate for each field line from a distribution with the same properties. There is no dependence of the heating rate on the loop length and every run draws from the same distribution. We make a further comment on this relationship with regard to our future plans for this research in Section~\ref{SandC}. The most probable heating rate in the distribution is $3-4\times10^7$~erg~cm$^{-2}$~s$^{-1}$ \citep[the active region energy loss rate is of order $10^7$~erg~cm$^{-2}$~s$^{-1}$:][]{Withbroe1977}.

The inter-event period was chosen to be proportional to the energy of the next event in the train \citep{Cargill2014}, where the constant of proportionality determines whether it is an intermediate- or a high-frequency train. The lower panel of Figure~\ref{fig2} shows the distribution of inter-event periods for the nanoflare train experiments. The intermediate frequency train has a most probable inter-event period of $\approx2000$~s, with very occasional delays of 10,000~s, and the high frequency train has most probable inter-event periods of $\approx500$~s, with occasional delays extending to 5000~s. Hence, during intermediate frequency trains the plasma on the strand is re-energized with inter-event periods generally comparable to the cooling timescale, allowing significant cooling and draining between heating events. During high frequency trains the plasma is re-energized with inter-event periods generally shorter than a cooling timescale. These timescales are consistent with previous estimates in the range $500-2000$~s \citep{Dahlburg2005,UgarteUrra2014,Cargill2014}. The individual nanoflares in each experiment all lasted for 120~s and followed a triangular temporal profile (a linear rise of 60~s to the peak heating rate, followed by a linear decay of 60~s). Each numerical experiment is run for three hours of solar time to allow at least one heating/cooling cycle for the longest field lines in the active region.

\begin{figure}
\includegraphics[width=\textwidth]{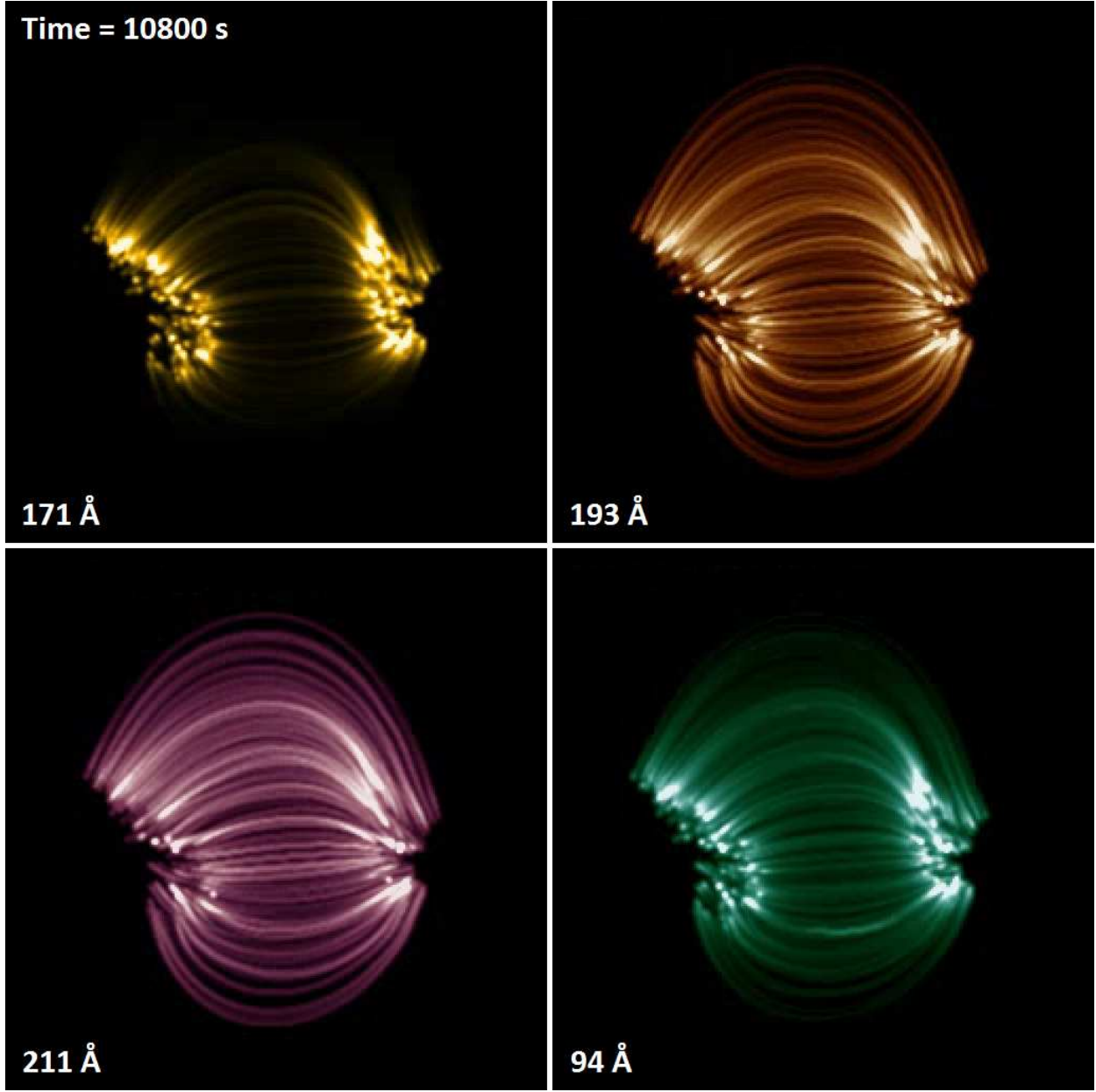}
\caption{Synthetic observations constructed for a selection of SDO/AIA channels in the intermediate frequency case at the conclusion of the three hours of simulated solar time.}
\label{fig3}
\end{figure}

Figure~\ref{fig3} shows a set of synthetic observations constructed for a selection of SDO/AIA channels. The observations show the distribution of emission across the model active region in the intermediate frequency case that would be observed by AIA, in each of the channels, at the conclusion of the three hours of simulated solar time. There are several steps to producing the observations. In the first step, the forward modeling software calculates the wavelength-integrated intensity (for a given AIA channel) as a function of position along each loop, at the spatial resolution of the hydrodynamic model. The emission along each loop is then rebinned as a function of position at the spatial resolution of AIA (0.6\arcsec pixels) and folded through the appropriate response function to convert to instrument units (DN~pixel$^{-1}$~s$^{-1}$). Consequently, there are $N_i$ emission values along the loop, where $i$ is the loop number. This is repeated at each time-step for which output from the hydrodynamic code is available (typically in 1~second intervals). The emission is summed over ten second intervals and then divided by five to emulate the $\sim2$~second exposures of AIA (this yields an average, rather than needing to choose which emission to throw-out).

Each loop has associated with it a set of $N_i$ 3D spatial coordinates ($x,y,z$) that trace out its magnetic field and so for each coordinate there is an emission value (DN~pixel$^{-1}$~s$^{-1}$). Therefore, the magnetic field data and the emission are both at the AIA spatial resolution. In the next step, the 3D field line coordinates for each loop are projected onto a 2D image array (assumed to be a plane oriented perpendicular to the line-of-sight) and the corresponding emission values are added to the image elements they intersect. We now have a 2D array, where each element corresponds to a single AIA pixel, populated with intensities that describe the pattern of emission across the model active region.

In the final step, the intensity at each pixel is convolved with a Gaussian to emulate the detector point-spread function. The Gaussian widths for AIA are from \cite{Grigis2012}, multiplied by a factor $2\sqrt{2\ln2}$ to obtain the FWHM required by the convolution function. The final product is a 2D intensity image, matching the spatial resolution of AIA, that can be treated exactly like real observational data and analyzed with the same diagnostic methods and tools, such as the time lag technique that we employ here. Furthermore, the intensity images can be scaled to the standard color table allocated to each AIA channel (accessed via the AIA\_LCT function in SolarSoft) to create synthetic observations (Figure~\ref{fig3}) in the manner of \cite{Warren2007} and \cite{Winebarger2011}. Movies can be constructed by repeating this procedure to create each frame from successive synthetic images.

\section{Results}
\label{results}

\begin{figure}
\includegraphics[width=\textwidth]{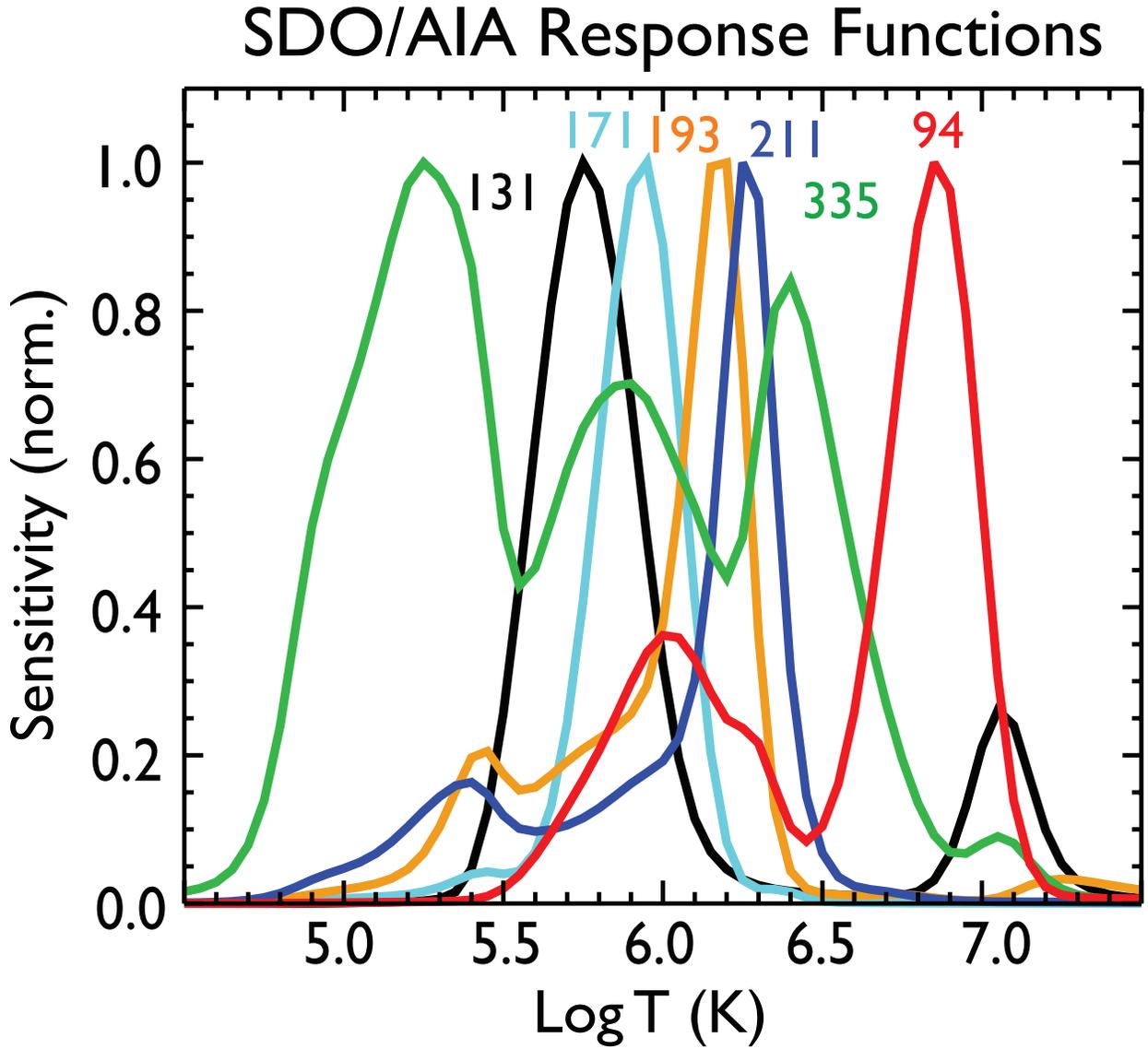}
\caption{The temperature dependent response functions for each of the AIA channels \citep[reproduced from][]{Viall2015}.}
\label{fig4}
\end{figure}

We applied the time lag technique to each numerical experiment and computed a time lag map for every pair of AIA channels (see Figure~\ref{fig4} for their response functions). The time lag method works by cross correlating light curves from two different AIA channels across a range of temporal offsets. We tested every temporal offset up to $\pm3600$~s. The time lag method identifies which of the two channels exhibit light curves with variable emission first (i.e. an intensity change), and the amount of time until the second one images the same variable emission. 
We focus on a representative subset of 6 of the 15 possible maps, chosen such that they span the full AIA temperature range. [The findings from all 15 maps for each of the experiments are summarized and quantified in the form of histograms in Figures~\ref{fig8} and \ref{fig9}, which are discussed at the end of this Section.]

\begin{figure}
\includegraphics[width=\textwidth]{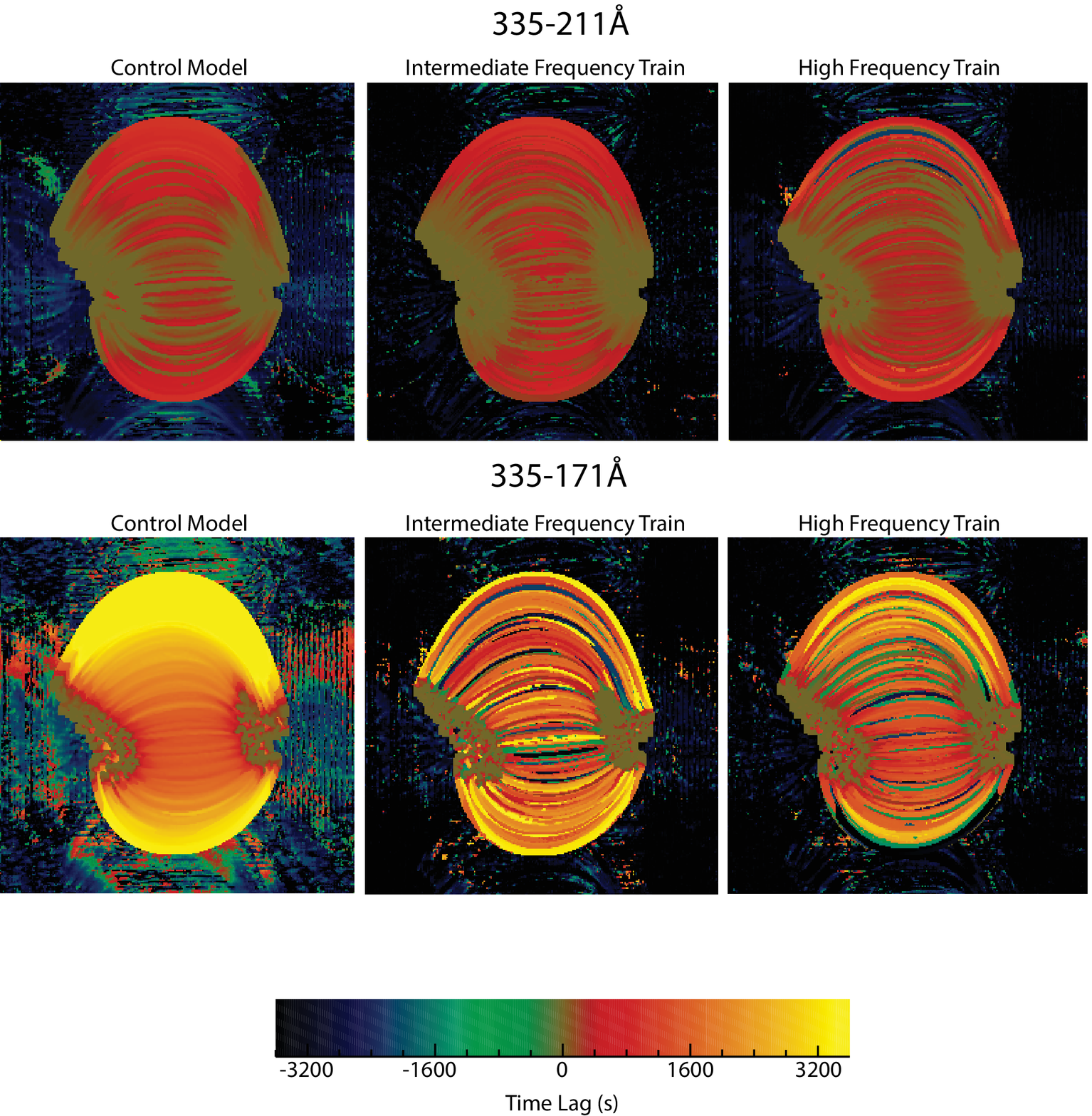}
\caption{Time lag maps for the $335-211$ and the $335-171$ ~\AA~pairs.}
\label{fig5}
\end{figure}

In Figure~\ref{fig5}, we show the time lag maps computed between the $335-211$~\AA~channel pair in the upper panels, and the $335-171$~\AA~pair in the lower panels. We show the results from the control experiment in the left-hand column, the intermediate frequency nanoflare train in the middle column, and the high frequency train in the right-hand column. On this color scale: reds, oranges and yellows indicate a positive time lag, which means that variable emission was present in the 335~\AA~channel first, and then either in the 211~\AA~(upper panels) or 171~\AA~(lower panels) channels; greens, blues and black indicate negative time lags, where variability occurred in the 211 or 171~\AA~channels first; and olive green indicates variability that occurs at nearly the same time in the two channels (within tens of seconds on the color scale shown here). Again, we emphasize that olive green does not indicate a lack of correlation or a lack of variability; it indicates variability with zero (within the instrument cadence) or close to zero time lag.

\begin{figure}
\includegraphics[width=\textwidth]{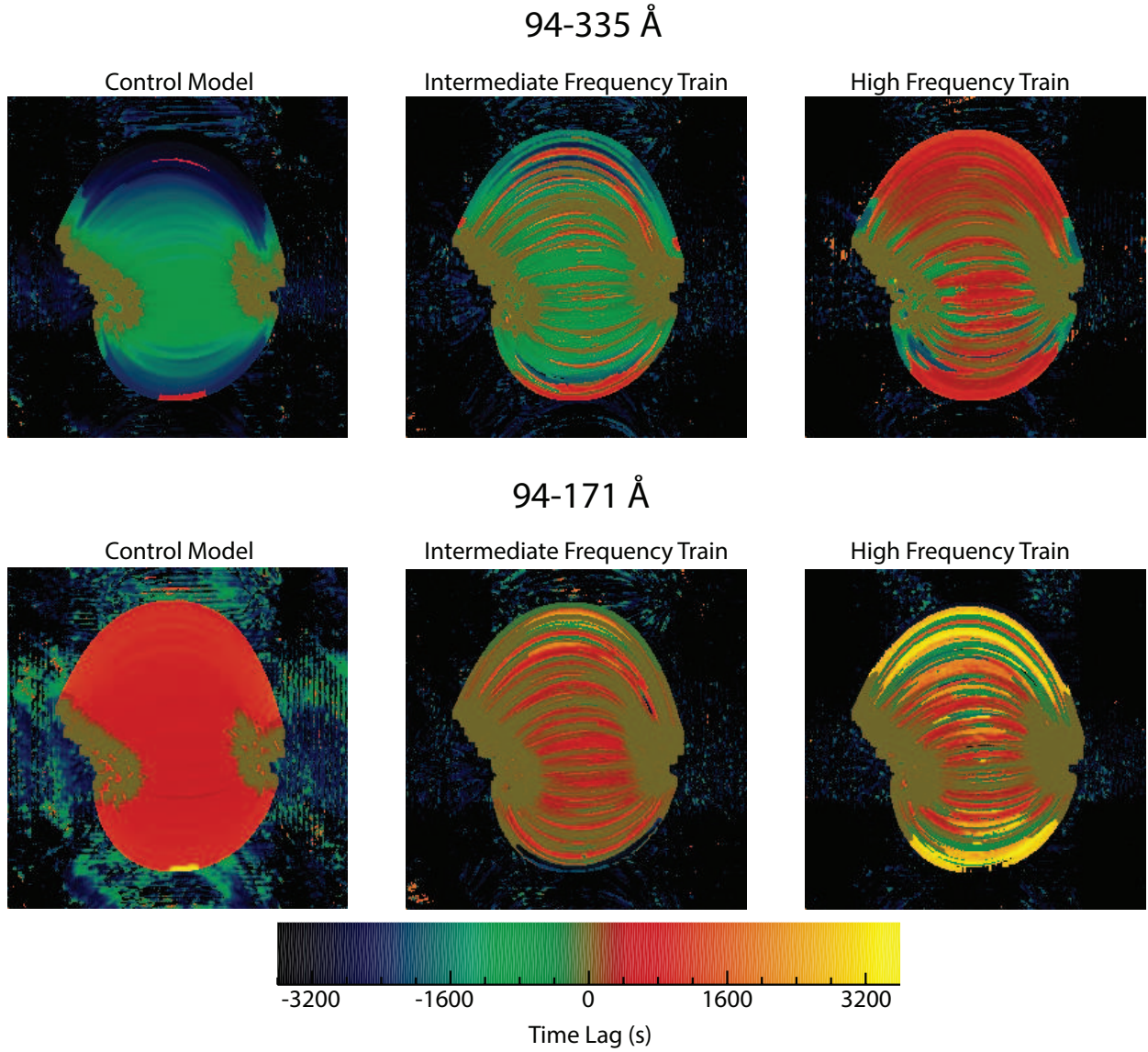}
\caption{Time lag maps for the $94-335$ and the $94-171$ ~\AA~pairs.}
\label{fig6}
\end{figure}

\begin{figure}
\includegraphics[width=\textwidth]{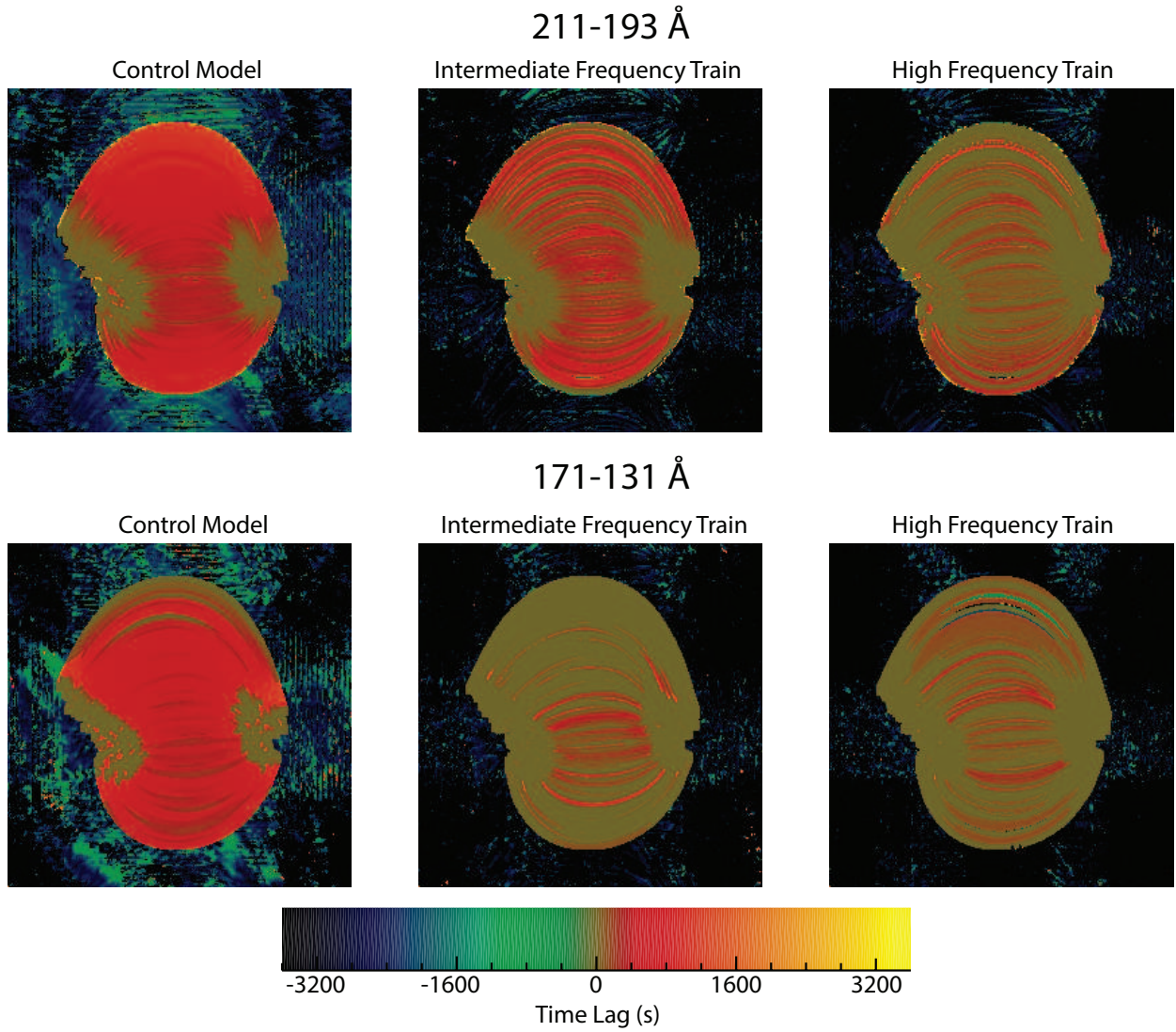}
\caption{Time lag maps for the $211-193$ and the $171-131$ ~\AA~pairs.}
\label{fig7}
\end{figure}

\begin{figure}
\includegraphics[width=\textwidth]{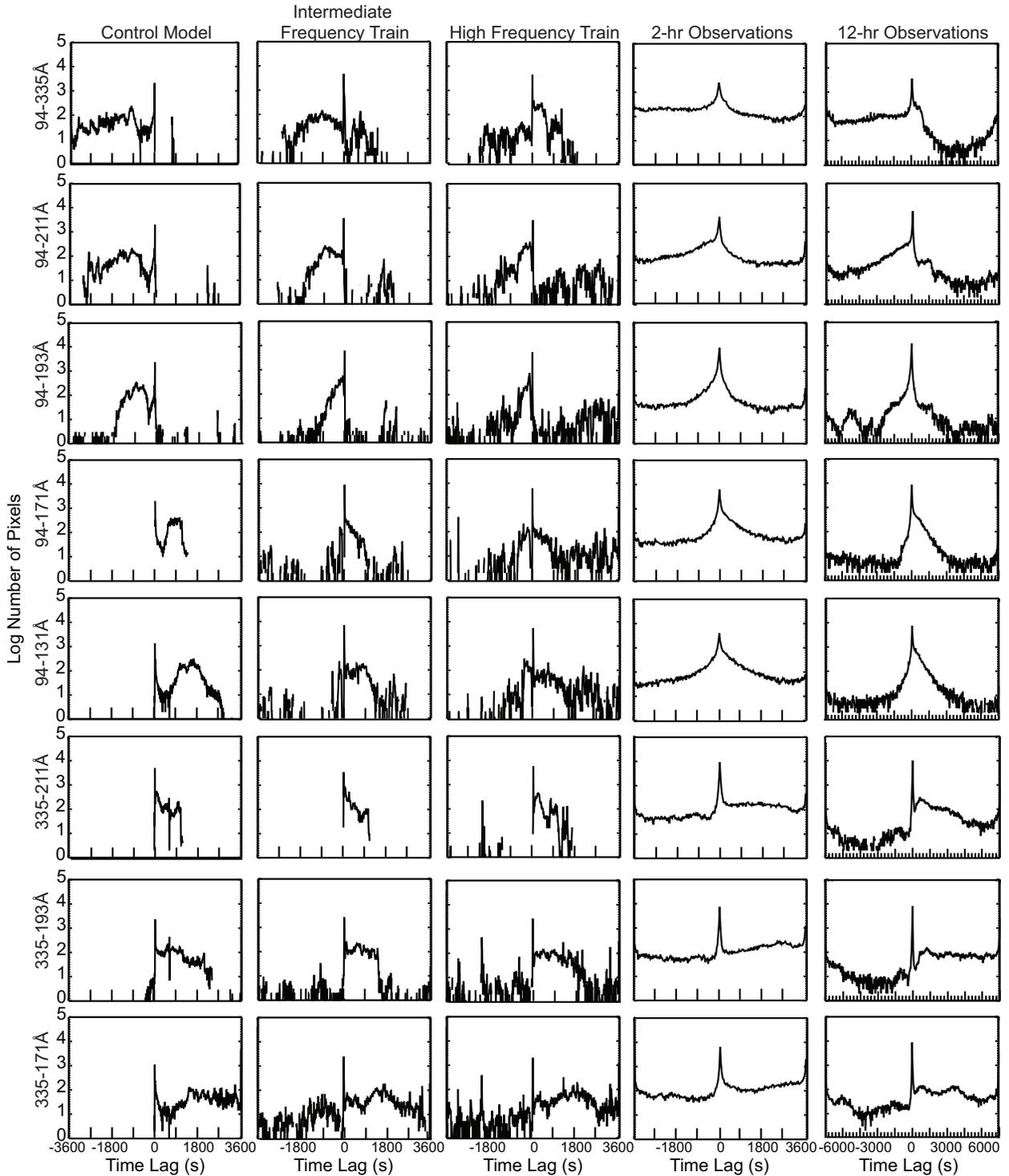}
\caption{Histograms of the time lags found for each of the models and two observations from \cite{Viall2012}. The channel pair is listed to the left and the Y-axis is the $\log_{10}$ of the number of pixels. The time lags for the models and the 2 hour observations span $\pm$3600 seconds. The time lags for the 12 hour observations span $\pm$7200 seconds.}
\label{fig8}
\end{figure}

\begin{figure}
\includegraphics[width=\textwidth]{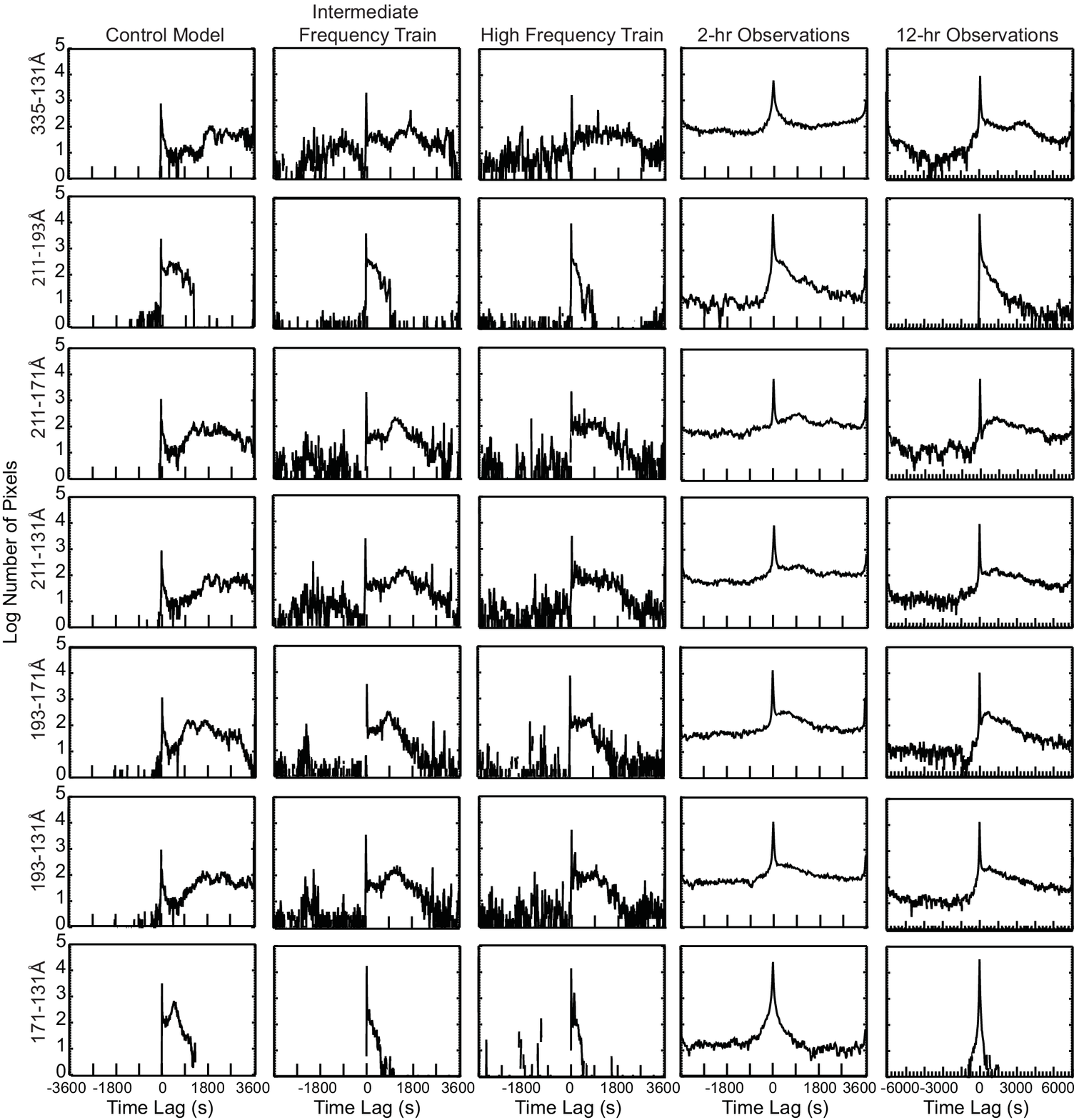}
\caption{See Figure~\ref{fig8}.}
\label{fig9}
\end{figure}

In the case of the control experiment, for which a single nanoflare is triggered on all field lines at $t=0$~s with no further events, all of the time lag maps are dominated exclusively by post-nanoflare cooling (see also the left-hand panel of Figures~\ref{fig6} and \ref{fig7}, which also exhibit exclusively post-nanoflare cooling). This is as expected, since the radiative signature from cooling magnetic strands dominates the light curves \citep{Bradshaw2011a}, with little contribution from heating strands because this phase is: (a) short-lived; and (b) the density is low. The cooling phase is well underway as the strands fill and brighten. Cooling remains the dominant feature for the time lag maps corresponding to the intermediate- and high-frequency nanoflare experiments too, though more structure begins to emerge due to the re-energization of the plasma on the magnetic strands, the effects of which we will discuss throughout the remainder of this section. 

Overall, the distribution of cooling times throughout the model active region is consistent with the majority of the observed time lags presented by \cite{Viall2012} (from NOAA AR 11082, observed on June 19 2010) for all channel pairs, especially those computed from the two-hour light curves (maximum timescales $\sim 1$~hour). For example, the time lags between the 335 and 211~\AA~channel pairs are shorter than between the 335 and 171~\AA~pairs, as expected since 171~\AA~is sensitive to cooler plasma than 211~\AA. This can also be seen in the corresponding histograms (Figure~\ref{fig8}) where the distribution of time lags broadens with increasing channel temperature separation ($335-211$, $335-193$, $335-171$~\AA).

Particular patterns seen in the observational data are also reproduced by the model. For example, the shorter flux tubes in the core of the model active region generally cool more quickly than the longer loops at the periphery. This is most easily seen in the $335-171$~\AA~pair, though it is evident in all of the time lag maps. This can be compared with Figure~7 of \cite{Viall2012}, where it is clear that the short loops in the active region core have time lags corresponding to cooling times of $\leq 1800$~s between channel pairs, whereas the long loops towards the edges have time lags approaching one hour. Note that the color/time scales are the same as the two-hour maps of \cite{Viall2012} and so the model active region exhibits both qualitative and quantitative agreement with the observations, though it is not our aim to reproduce any particular set of observations here but to demonstrate consistency with the broad properties and characteristics of activity patterns.

One of the key findings presented by \cite{Viall2012} was the presence of zero time lags (within the uncertainty of the instrument cadence) for every channel pair at loop foot-points and within moss regions. They interpreted this as due to the transition region response to coronal nanoflares. Here we are using the temperature-indepedent definition of the transition region introduced by Vesecky et al. (1979), where the top of the transition region is the physical location where thermal conduction changes from an energy sink (above) to an energy source (below). Using this physical definition, the transition region reaches a temperature of roughly 60\% of the peak coronal temperature in a flux tube (near hydrostatic equilibrium, at the onset of radiative cooling). Since coronal nanoflares produce peak temperatures greater than 5~MK, the transition region can extend to 3 MK and higher, and emits throughout the temperature ranges that the six AIA channels are sensitive to. Therefore, one cannot assume that all of the emission observed in any given AIA channel is coronal when it can clearly be dominated by transition region emission along lines-of-sight to loop foot-points and moss regions.

\cite{Viall2015} used the EBTEL code \citep{Klimchuk2008,Cargill2012} to show that zero time lags in transition region plasma are expected when the corona is heated in-situ by nanoflares. This is because all layers (temperatures) of the transition region respond in unison. The emission in different channels brightens and fades together with a near zero temporal offset, or time lag, between light curves. This is true for light curves from any instrument that images transition region plasma. \cite{Viall2015} provided additional proof that zero time lags were caused by transition region emission by observing an off-limb active region and showing that well above the transition region far fewer zero time lags are measured. The time lag maps for all channel pairs produced in the current work, for all three numerical experiments, are consistent with these earlier findings. All 45 maps (3 simulations $\times$ 15 AIA channel combinations) show zero time lags at the loop foot-points (Figures~\ref{fig5} to \ref{fig7}, for example). This is also evident in each of the model-generated histograms (Figures~\ref{fig8} and ~\ref{fig9}) where all of the distributions peak at a time lag of zero seconds, consistent with the histograms derived from observational data.


The 94~\AA~temperature response function of AIA has a bimodal distribution, with a strong peak due to Fe~XVIII emission at 7~MK and a weaker ($\sim30\%$) peak near 1~MK. The nature of this distribution makes time lag maps involving this channel less straightforward to interpret and more challenging to reproduce. Figure~\ref{fig6} shows the $94-335$ and $94-171$ ~\AA~channel pairs for each of the three numerical experiments. The hot component of the 94~\AA~channel lies at a higher temperature than the 335~\AA~channel, whereas the warm component of the 94~\AA~channel falls below it, but above the temperature of the 171~\AA~channel. The $94-335$~\AA~time lag map for the control/cooling experiment in Figure~\ref{fig6} shows almost exclusively negative time lags (with the exception of the zero time lags at the loop foot-points) indicating that the 335~\AA~emission changes before the 94~\AA~emission. We know for certain that the plasma is cooling because it is not re-energized in this experiment. Therefore, the dominant signal in this channel pair must be due to cooling from the 335~\AA~channel to the warm component of the 94~\AA~channel (2.5 to 1.1~MK). Emission in the 94~\AA~channel is consequently dominated by 1.1~MK plasma.

Consistent with this picture, the $94-171$ and $335-171$~\AA~channel pairs show exclusively positive time lags outside the loop foot-points in the control experiment (Figures~\ref{fig5} and \ref{fig6}). Both components of the 94~\AA~channel lie above the temperature sensitivity of the 171~\AA~channel. The control/cooling and intermediate frequency nanoflare train experiments show time lags that are predominantly $<1800$~s because the cooling is from the warm component of the 94~\AA~channel (1.1~MK, as we have seen above), which is only slightly warmer than the 171~\AA~channel (0.9~MK). The high frequency nanoflare train experiment exhibits some longer time lags, particularly in the longer loops, where the cooling is from the hot component of the 94~\AA~channel (7 to 0.9~MK). We note that it is more likely the plasma will be re-energized in longer loops, before it reaches the cool component of the 94~\AA~channel, due to their extended cooling timescales. By the same token, it is also clear from the control experiment that the shorter loops within the core cool more quickly, with shorter time lags, than the longer loops at the periphery.

The intermediate frequency nanoflare train experiment shows mostly negative time lags in the $94-335$~\AA~map, but also zero and some positive time lags. The time lags are constrained to a range of $\sim \pm1800$~s, whereas for the control experiment the negative time lags for the longer loops significantly exceed 1800~s. This shows that the loops are being prevented from cooling for extended time periods, as expected from a re-energization scenario. The high frequency train experiment shows predominantly positive time lags, constrained below 1800~s. In both the intermediate- and high-frequency experiments, positive time lags in the $94-335$~\AA~maps indicate flux tubes where 7~MK emission dominates the 94~\AA~channel.

Within the context of intermediate-/high-frequency nanoflare trains, the reason for these patterns of evolution lies in the fact that only occasionally in the intermediate frequency case is sufficiently high density and temperature plasma created for the hot component of the 94~\AA~channel to determine the time lag. More often, the emission measure peaks below the sensitivity range of the hot component of the 94~\AA~channel and the 335~\AA~channel is detected first, yielding a negative time lag. This happens, for example, if the corona has substantially cooled and drained before the onset of the next nanoflare. The rapidly heated corona then has a very low density, which leads to fast cooling by thermal conduction. By the time sufficient material is ablated, the temperature may be well below the hot peak of the 94~\AA~channel. The hot component of the 94~\AA~channel determines the time lag of the channel pair more often in the high frequency case because the corona is maintained at an overall higher density and temperature throughout the experiment, so each heating event is more likely to take the plasma into the temperature range of the hot component with sufficient emission for it to dominate. This example serves to highlight the potential importance of a distribution of event frequencies in reproducing the observed activity patterns. High frequencies are required to create the dense plasma that can produce a bright, hot component in the 94~\AA~channel, but then observations show there must be time for the plasma to cool into the 171~\AA~channel, which is a lower frequency situation. Therefore, defining nanoflare heating within a narrow space of frequencies (e.g. exclusively high- or low-frequency, etc.) is likely to be unsatisfactory in terms of understanding and explaining active region properties and evolution. This conclusion is reinforced by the histograms shown in Figure~\ref{fig8} corresponding to the pairs that include the 94~\AA~channel, which show that even high frequency heating can yield both negative and positive time lags.

Comparing the model-predicted time lag maps in Figure~\ref{fig6} with the observationally measured maps presented in Figure~6 of \cite{Viall2012} we note several encouraging similarities. The time lag maps predicted by the intermediate frequency experiment show predominantly negative time lags for the $94-335$~\AA~channel pair, also evident in the corresponding histogram, as seen in the observed maps. In contrast, the high frequency experiment predicts mostly positive time lags of which there are occurences only in the very core of the observed active region \citep[in the region of pixel location 200,200 in Figure~6 of][]{Viall2012}. Two possible explanations are that: (a) active region cores tend to be hotter because the magnetic field strength is greater and so more magnetic energy is available for heating, and cooling then takes place from the hot component of the 94~\AA~channel; or (b) the stronger magnetic fields and faster Alfv\'{e}n speeds lead to shorter re-energization timescales, so the repeat time on the shorter flux tubes found in active region cores is shorter and the plasma can be maintained at densities and temperatures consistent with the hot component of the 94~\AA~channel. Here we note that if one defines the inter-event period relative to the cooling timescale then the event frequency increases with loop length \citep{LopezFuentes2015}; since the cooling timescale for longer loops is increased (in proportion to $L^2$) then even long inter-event periods may correspond to high frequency heating. Hence the particular definition of the event frequency, whether it is defined in absolute terms or relative to some other physical timescale of the system, is important.

We also note the presence of zero time lags along the coronal portion of some of the flux tubes in the two nanoflare train experiments. This feature is seen in the 94~\AA~channel pairs, but is particularly evident in the model-predicted time lag maps for the $211-193$ and $171-131$~\AA~channel pairs in Figure~\ref{fig7}. Figure~6 of \cite{Viall2012} also shows large regions of zero time lag in the corona for the same channel pairs. The reason for this pattern of activity in the case of the $211-193$~\AA~pair is that these channels are quite close together in temperature sensitivity and exhibit significant overlap \citep[e.g. Figure~2 of][]{Viall2015}. The $171-131$~\AA~pair are the two coolest channels and the cooling phase is typically interrupted before the plasma fully passes through them. This is shown by contrasting with the control/cooling experiment in which the loops are allowed to fully cool through these two channels (only the very longest loops don't quite have sufficient time), and demonstrates the value of having such an experiment to serve as a basis for interpreting the results from more complicated heating scenarios.

To summarize and quantify these findings we have constructed histograms to show the distribution of time lags for every map calculated for each run/channel pair and also for the 2 and 12 hour time windows analyzed by \cite{Viall2012} to serve as a basis for comparison. In Figure~\ref{fig8} we show histograms for 8 of the channel pairs: $94-335$; $94-211$; $94-193$; $94-171$; $94-131$; $335-211$; $335-193$; and $335-171$~\AA. In Figure~\ref{fig9} we show the histograms for the remaining 7 channel pairs: $335-131$; $211-193$; $211-171$; $211-131$; $193-171$; $193-131$; and $171-131$~\AA. In the first three columns we show the results computed for the control experiment, the intermediate frequency nanoflare experiment and the high frequency nanoflare experiment, respectively. In the last two columns we reproduce results for one of the 2 hour time windows ($00:00-02:00$~UT) and for one of the 12 hour time windows ($00:00-12:00$~UT) analyzed by \cite{Viall2012}. The Y-axis for all histograms is on a logarithmic scale. The total number of pixels in the observations were slightly fewer than in the model and so we normalized the histograms corresponding to the observations to the model histograms. This was done by dividing by the total number of pixels in the observational image and then multiplying by the total number of pixels in the model image such that they could be directly compared on the same scale. The three models and the 2 hour observation were all tested with an identical time lag range of $\pm3600$~seconds. Note that the 12 hour observations were tested with a range of $\pm7200$~seconds.

To construct the histograms we only included pixels from the model images that contributed non-zero emission and we only included pixels from the observed images that contained the active region core and corresponding foot-points. We excluded pixels associated with the quiet sun surrounding the active region as well as those associated with fan loops, as neither are part of the model. This is in contrast to the histograms shown in \cite{Viall2012}, which included the results from the full field of view. In practice, we found that using the 335~\AA~ images with a threshold of 10~DN~s$^{-1}$ in each pixel worked well to capture the active region core and transition region foot-points, while excluding the fan loops and the surrounding quiet sun. All pixels within this area are included in the histograms.

Taking all of the time lag maps and the results shown in the histograms together, it is clear that the control experiment of pure cooling (after initial energization of the plasma) predicts time lag maps that disagree with observations in fundamental ways, providing strong evidence that re-energization of the plasma on a magnetic strand must occur and the inter-event period must be long enough to allow cooling, but is generally comparable to a cooling timescale. Nonetheless, full cooling is observed in some areas of the active region. There is some evidence that intermediate frequency heating can reproduce some features of the observed histograms more closely than high frequency heating. For example, the $94-171$ and $94-131$~\AA~channel pairs have short, negative time lags, due to rapid re-energization, in the high frequency case that are not a prominent feature of intermediate frequency heating and are not present in the observed results. Furthermore, the $211-171$ and $193-171$~\AA~pairs, and to a somewhat lesser extent the $211-131$ and $193-131$~\AA~pairs, have a prominent/most-probable 'bump' in the observed and intermediate frequency histograms, in the region of $800-1500$~s, that does not appear in the high frequency case. The location and width of these bumps may prove to be a useful diagnostic of the characteristic re-energization timescale in an active region and even to compare timescales across an active region.

A mixture of statistically intermediate and high frequency trains, with most probable inter-event periods of $\approx500$~and~2000~s, and occasional delays extending to 10,000~s, can reproduce much of the observed structure in the time lag maps across the active region. This could doubtless be fine-tuned to a particular active region to achieve a close match between model predictions and observations but, as we stated, our intention here is to capture the broad properties and characteristics of the active region activity patterns. Furthermore, the distribution of inter-event frequencies (and volumetric heating rates) could be determined differently and depend instead on the field strength and/or loop length, for example. We discussed this possibility above and will return to it in Section~\ref{SandC}. In summary, comparing the model- and observation-derived distributions shows it is likely that a range of heating frequencies are operating across active regions. However, the particular definition that one uses for the event frequency is important in this context.

\section{Summary and Conclusions}
\label{SandC}

We have performed several numerical experiments to investigate the patterns of activity observed in the solar corona. By combining a hydrodynamic and forward modeling code with a magnetic field extrapolation we have created a model active region that can be subject to different nanoflare heating scenarios. By using visualization techniques we can produce synthetic observations (images and movies) that can be analyzed in the same way as real observational data. In the first experiment we triggered a nanoflare on all of the magnetic strands comprising the active region at $t=0$~s and simply allowed the plasma to cool, with no further events thereafter. In the next two experiments we subjected each magnetic strand to statistically intermediate- and high-frequency nanoflare trains, respectively. During intermediate frequency trains the plasma on the strand was re-energized with inter-event periods that were generally comparable to the cooling timescale, allowing significant cooling and draining between heating events. During high frequency trains the plasma was re-energized with inter-event periods that were generally shorter than a cooling timescale. The inter-event period was made a function of the energy of the next event in the train, which was drawn from a power-law with a slope of -2.5 to favor the selection of lower energy nanoflares. A method for measuring the dominant time lag between emission appearing in pairs of AIA channels was applied to the synthetic observational images generated from the numerical results. In general, cooling is expected to produce the dominant signal because heating is short-lived and energizes lower density plasma than the plasma cooling and draining from the corona. We compared our model-predicted time lag maps with a set of observationally measured maps. We emphasize that our aim here was not to reproduce a particular set of observations but to examine the consistency of the nanoflare heating scenario with the broad properties and characteristics of the activity patterns observed in the time lag maps of real coronal structures. Our major conclusions are:

1. The time lag method clearly identifies the cooling timescales between the channel pairs throughout the model active region and they are generally consistent with those observed.

2. Shorter flux tubes in the core of the model active region are shown to cool more quickly than the longer loops at the periphery, in common with observations.

3. The time lag maps produced by all three numerical experiments, for all channel pairs, show zero time lags for lines-of-sight through the loop foot-points, due to transition region dynamics. This is in agreement with observations showing zero time lags at the foot-points of coronal loops and in moss regions (providing yet more evidence for moss as the foot-points of hot, overlying loops).

4. The model can reproduce the activity patterns observed using particular channel pairs, including those with more complicated bimodal temperature responses (e.g. the $94-335$ and $94-171$~\AA~channel pairs). The control/cooling experiment disagrees with observations in fundamental ways, particularly with regard to the $171-131$ and $94-335$~\AA~channel pairs, providing strong evidence that the plasma must be re-energized on timescales that are comparable to the cooling timescale to reproduce the coronal structures. The time lag maps predicted by the intermediate frequency experiment show predominantly negative time lags for the $94-335$~\AA~channel pair, more consistent with what is observed, but the high frequency experiment predicts mostly positive time lags that were seen in the observed active region core. Hence, it is likely that a relatively broad spectrum of heating frequencies are operating across active regions.

5. The model-predicted time lag maps in the $211-193$ and $171-131$ ~\AA~channel pairs show a substantial number of zero time lags along entire flux tubes that are also seen between the same channel pairs in observed active regions, providing additional evidence that a necessary property of the heating mechanism is plasma re-energization before the flux tube cools below $1-2$~MK and drains.

The methodology for global active region modeling introduced here opens many avenues to further exploration. For example, in the time lag maps produced using the $94-335$, $94-211$, and $94-193$~\AA~channel pairs presented by \cite{Viall2012}, postive time lags (cooling from the hot component of the 94~\AA~channel) were seen in the very core of the active region, whereas negative time lags were predominantly seen elsewhere. Why should this be so? We suggested above that the greater magnetic field strength leads to hotter active region cores either because more energy is released in each heating event or shorter re-energization timescales maintain the plasma at a higher density and temperature. We will explore these possibilities based on a set of NLFFF extrapolations (H. Warren, private communication) corresponding to the active regions surveyed by \cite{Warren2012}, upon which the time lag analysis has already been performed, to determine the individual contributions of the amount energy released and the inter-event period (both connected to the properties of the magnetic field provided by the extrapolation) to establishing the observed activity patterns.

In addition, we also plan to extend the physics treatment of the high temperature, low density plasma following nanoflare heating, by treating both the strong heat flux limiting (when the supported flux is decoupled from the local temperature and temperature gradient, and in the presence of ion acoustic turbulence) and non-local contributions to the heat flux \citep{Gray1980,Luciani1983,Luciani1985,Karpen1987,Bradshaw2006,West2008}. We will also run global active region models for non-equilibrium ion populations and calculate the corresponding time lag maps for comparison with the corresponding ionization equilibrium cases, which may be important in treating the hot components of channels with bimodal temperature responses when there is a significant delay in creating the charge states. Strong heat flux limiting may keep the corona at a high temperature for longer, perhaps allowing the charge states more time to catch up with the thermodynamic conditions. We will investigate the interplay between these physical processes. Finally, we will also compare the global active region model predictions with other observables such as Doppler velocity maps and emission line widths, to gain additional insights and perspectives on the evolution and activity patterns within solar active regions subject to nanoflare heating.

\acknowledgements
SJB is grateful to the NSF for supporting this work through CAREER award AGS-1450230 and NMV to NASA through the GI program. The authors benefited from participating in the team hosted by the International Space Science Institute, Bern, on Using Observables to Settle the Question of Steady versus Impulsive Coronal Heating, led by SJB and Helen Mason. The authors also thank Dr. Jim Klimchuk and the referee for their helpful comments on the manuscript.


\end{document}